\documentclass[11pt,a4paper]{article}
\usepackage{jcappub}
\usepackage{verbatim}
\usepackage{cleveref}

\usepackage[T1]{fontenc}
\usepackage[utf8]{inputenc}
\usepackage[american]{babel}
\usepackage{epsfig}
\usepackage{graphicx}
\usepackage{booktabs}
\usepackage{multirow}
\usepackage{dcolumn}
\usepackage{amsmath}
\usepackage{mathtools}
\usepackage{amsfonts}
\usepackage{amssymb}
\usepackage{epstopdf}
\usepackage{bm}
\usepackage{siunitx}
\usepackage{braket}
\usepackage{enumitem}
\usepackage{soul}
\usepackage[table]{xcolor}
\usepackage{color}
\usepackage{transparent}
\usepackage{comment}
\usepackage{pifont}
\usepackage{soul}

\definecolor{navyblue}{rgb}{0.0, 0.0, 0.5}
\definecolor{royalblue}{rgb}{0.25, 0.41, 0.88}
\definecolor{cadmiumgreen}{rgb}{0.0, 0.42, 0.24}
\definecolor{blue-violet}{rgb}{0.54, 0.17, 0.89}
\definecolor{darkviolet}{rgb}{0.58, 0.0, 0.83}
\definecolor{orange(colorwheel)}{rgb}{1.0, 0.5, 0.0}
\definecolor{burgundy}{rgb}{0.5, 0.0, 0.13}

\usepackage{hyperref}
\hypersetup{
    colorlinks=true, 
    linkcolor=royalblue, 
    citecolor=magenta}

\newcommand\ee{\end{equation}}
\newcommand\be{\begin{equation}}
\newcommand\eea{\end{eqnarray}}
\newcommand\bea{\begin{eqnarray}}

\newcommand\eq[1]{Eq.~\eqref{eq:#1}}

\usepackage{booktabs}
\usepackage{multirow}
\usepackage{dcolumn}
\usepackage{colortbl}

\definecolor{magenta(process)}{rgb}{1.0, 0.0, 0.56}

\definecolor{darkspringgreen}{rgb}{0.09, 0.45, 0.27}

\definecolor{royalblue(web)}{rgb}{0.25, 0.41, 0.88}

\begin{document}

\title{A double take on early and interacting dark energy from JWST}

\author[a,b,1]{Matteo Forconi, \note{Corresponding author.}}
\author[b]{William Giar\`e,}
\author[c]{Olga Mena,}
\author[a]{Ruchika,}
\author[b]{Eleonora Di Valentino,}
\author[a]{Alessandro Melchiorri,}
\author[d,e]{Rafael C. Nunes}

\affiliation[a]{Physics Department and INFN, Universit\`a di Roma ``La Sapienza'', Ple Aldo Moro 2, 00185, Rome, Italy}
\affiliation[b]{School of Mathematics and Statistics, University of Sheffield, Hounsfield Road, Sheffield S3 7RH, United Kingdom}
\affiliation[c]{Instituto de F{\'\i}sica Corpuscular  (CSIC-Universitat de Val{\`e}ncia), E-46980 Paterna, Spain}
\affiliation[d]{Instituto de F\'{i}sica, Universidade Federal do Rio Grande do Sul, 91501-970 Porto Alegre RS, Brazil}
\affiliation[e]{Divis\~{a}o de Astrof\'{i}sica, Instituto Nacional de Pesquisas Espaciais, Avenida dos Astronautas 1758, S\~{a}o Jos\'{e} dos Campos, 12227-010, S\~{a}o Paulo, Brazil}

\emailAdd{matteo.forconi@roma1.infn.it}
\emailAdd{w.giare@sheffield.ac.uk}
\emailAdd{omena@ific.uv.es}
\emailAdd{ruchika.ruchika@roma1.infn.it}
\emailAdd{e.divalentino@sheffield.ac.uk}
\emailAdd{alessandro.melchiorri@roma1.infn.it}
\emailAdd{rafadcnunes@gmail.com}

\date{\today}

\abstract{The very first light captured by the James Webb Space Telescope (JWST) revealed a population of galaxies at very high redshifts more massive than expected in the canonical $\Lambda$CDM model of structure formation. Barring, among others, a systematic origin of the issue, in this paper, we test alternative cosmological perturbation histories. We argue that models with a larger matter component $\Omega_m$ and/or a larger scalar spectral index $n_s$ can substantially improve the fit to JWST measurements. In this regard, phenomenological extensions related to the dark energy sector of the theory are appealing alternatives, with Early Dark Energy emerging as an excellent candidate to explain (at least in part) the unexpected JWST preference for larger stellar mass densities. Conversely, Interacting Dark Energy models, despite producing higher values of matter clustering parameters such as $\sigma_8$, are generally disfavored by JWST measurements. This is due to the energy-momentum flow from the dark matter to the dark energy sector, implying a smaller matter energy density. Upcoming observations may either strengthen the evidence or falsify some of these appealing phenomenological alternatives to the simplest $\Lambda$CDM picture.}

\maketitle
\section{Introduction}

The minimal $\Lambda$CDM model of cosmology, described by only six fundamental parameters, has successfully explained a large number of cosmological observations at different scales. Nevertheless, in recent years, several anomalies have emerged, challenging the model across all cosmological epochs.

Currently, the most significant problem -- known as the Hubble tension~\cite{Verde:2019ivm,DiValentino:2020zio,DiValentino:2021izs} -- is represented by a mismatch at the level of $\sim 5\sigma$ between the value of the present-day expansion rate of the Universe ($H_0$) inferred from Planck-2018 CMB observations ($H_0=67.4\pm0.5$ km s$^{-1}$ Mpc$^{-1}$\cite{Planck:2018vyg}) and the value of the same parameter measured directly from local distance ladder measurements using Type-Ia Supernovae calibrated with Cepheid variable stars ($H_0=73\pm1$ km s$^{-1}$ Mpc$^{-1}$\cite{Riess:2021jrx}).

Another conundrum, albeit less significant than the Hubble tension, is the so-called $S_8$ tension. This parameter is closely related to the clustering of matter in the Universe, and its value can be inferred both from the measurements of CMB anisotropies as those from Planck, and -- more directly --from the measurements of galaxy lensing made by experiments such as the Dark Energy Survey (DES) \cite{DES:2021vln,DES:2022ygi}, and the Kilo-Degree Survey (KiDS) \cite{Heymans:2020gsg,KiDS:2020ghu}. While Planck data favors larger values of $S_8$, KiDS and DES seem to prefer lower ones, leading to a well-documented discrepancy ranging between $2$ and $3$ standard deviations~\cite{DiValentino:2020vvd}. \footnote{Very recently the actual disagreement between these experiments has been the subject of careful reevaluation, see, e.g., Ref.~\cite{Kilo-DegreeSurvey:2023gfr}.}

Yet another emerging anomaly -- which is the focal point of this study -- pertains to the observations recently released by the James Webb Space Telescope (JWST) of galaxies at very high redshifts (much higher than those commonly explored by large-scale structure galaxy surveys).
Interestingly, the preliminary JWST results point towards a population of surprisingly massive galaxy candidates (see e.g.~\cite{Santini_GLASS_Mstar,Castellano_GLASS_hiz,Finkelstein_CEERS_I1,Naidu_UVLF_2022,Treu:2022iti,Harikane_UVLFs,PerezGonzalez_CEERS, 2023ApJ...949L..18P}) with stellar masses of the order of $M \ge 10^{10.5} M_{\odot}$. In a recent study~\cite{2023Natu}, it was pointed out that the JWST data indicates a higher cumulative stellar mass density in the redshift range $7 < z < 11$ than predicted by the $\Lambda$CDM model, questioning one more time the canonical cosmological picture~\cite{Boylan-Kolchin:2022kae,Wang:2023xmm,Sun:2023ocn}. 

Despite the fact that recent comparisons of photometric and spectroscopic redshifts in overlapping samples of galaxies solidify the evidence for a high space density of bright galaxies at $z\gtrsim 8$ compared to theoretical model predictions~\cite{2023ApJ...951L..22A}, the JWST anomalous observations could still hint a lack of accuracy in extracting the intrinsic galaxy properties~\cite{Bouwens:2022gqg,xiao2023massive,2023arXiv231015284V,2023arXiv231107483W,2024ApJ...961...37C,Desprez:2023pif,Qin:2023rtf}. Other possibilities recently explored include the hypothesis of unusual dense regions due to the currently limited JWST observations (which only cover an area of approximately $38$ square arcminutes), as well as the possible presence of unknown systematics in CMB Planck polarization data~\cite{Forconi:2023izg,Giare:2023ejv}.

Although it is certainly premature to draw any definitive conclusions from these preliminary observations, if neither of the aforementioned possibilities can account for the discrepancies between the JWST results and the theoretical predictions of a baseline $\Lambda$CDM cosmology, it may be necessary to consider modifications to the model itself~\cite{Adil:2023ara,Menci:2022wia,Wang:2022jvx,Gong:2022qjx,Zhitnitsky:2023znn,Dayal:2023nwi,Maio:2022lzg,Domenech:2023afs,Liu:2022bvr,Yuan:2023bvh,Hutsi:2022fzw,Jiao:2023wcn,Biagetti:2022ode,Menci:2024rbq,Parashari:2023cui,Lovell:2,Haslbauer:2022vnq,Gandolfi:2022bcm,Lovyagin:2022kxl,Wang:2023ros,Hirano:2023auh,Paraskevas:2023itu} or to the galaxy formation process~\cite{Qin:2023rtf,Pallottini:2023yqg,Ferrara:2022dqw,Pacucci:2023oci,2023arXiv231105030M}. In this work, we take a step forward in this direction by testing alternative models where galaxy evolution could be notably different than in $\Lambda$CDM. In particular, we consider extensions related to the dark energy sector of the theory as possible phenomenological alternatives to explain the JWST preliminary findings. We test Early Dark Energy (EDE) and Interacting Dark Energy (IDE) cosmologies as both these extended scenarios, featuring modifications in the growth of structure, might predict a different evolution of perturbations, potentially resulting in the formation of more massive galaxies. We demonstrate that while EDE emerges as an excellent candidate to explain (at least partially) the unexpected JWST preference for larger stellar mass densities, IDE is generally disfavored by JWST measurements, despite yielding higher values of matter clustering parameters $\sigma_8$ and $S_8$.

The structure of the paper is as follows: in \autoref{sec:extended} we further elaborate on the reasons why we choose Early Dark Energy and Interacting Dark Energy scenarios and also we define the physics and the basic equations governing the cosmological evolution in these models. In \autoref{sec:methodology} we describe the main observables considered in this work, namely, the cumulative stellar mass density, the observational data and the likelihoods exploited in the numerical analyses. \autoref{sec:results} contains our main results. We conclude in \autoref{sec:conclusions}.

\begin{figure*}[htp]
\centering
\includegraphics[width=0.86\textwidth]{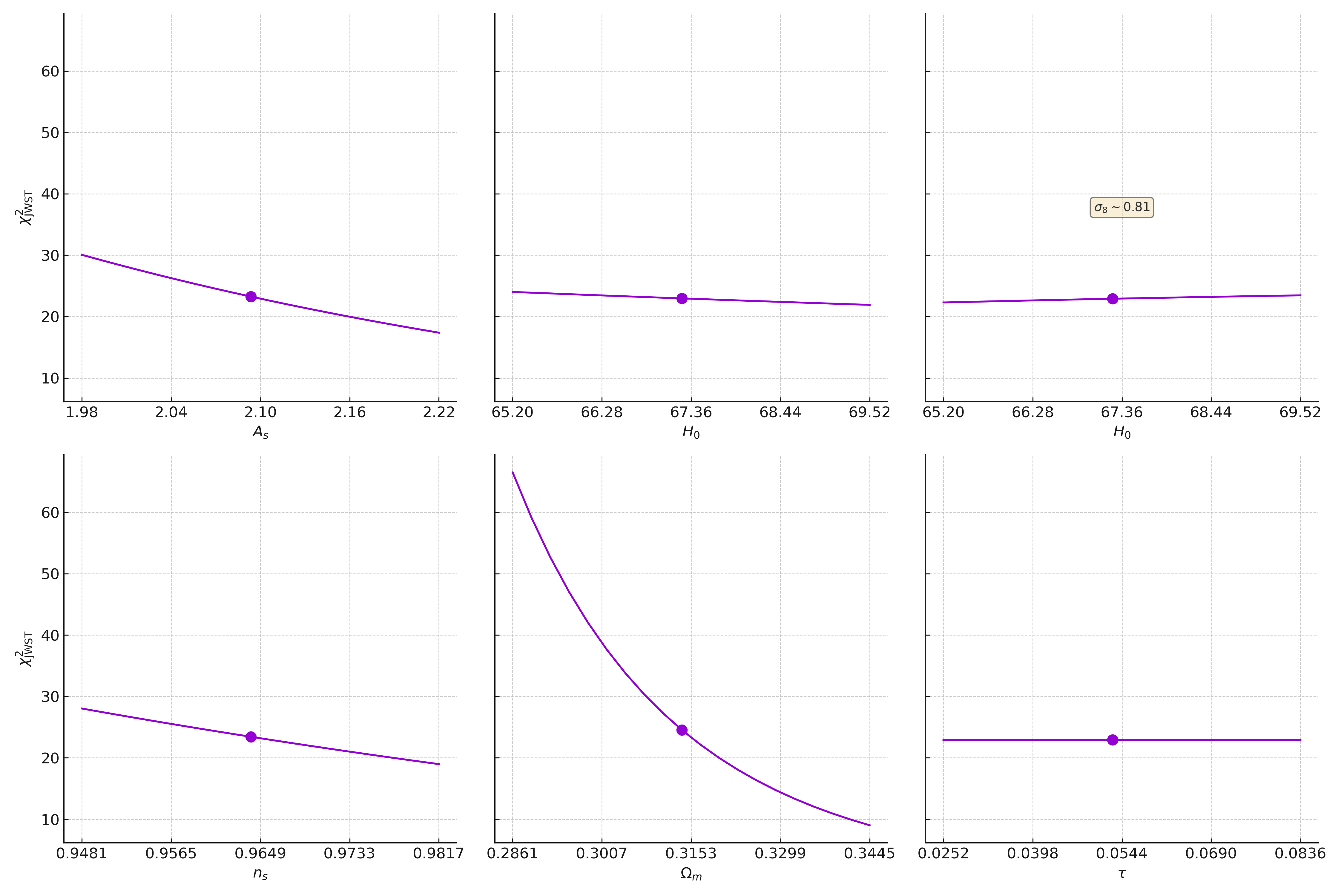}
\caption{Changes in $\chi^2_{\rm JWST}$ when varying a single parameter while keeping the others fixed to the Planck $\Lambda$CDM best-fit values (bold points in the figure): $\Omega_bh^2=0.02238$, $\Omega_ch^2=0.1201$, $H_0=67.32$ km/s/Mpc, $n_s=0.9659$, $A_s\times10^9=2.1$, $\tau=0.0543$, and $\Omega_mh^2=0.143$. In the top panel's third plot, when $H_0$ is free to vary, $\sigma_8$ is kept fixed by rescaling $A_s$ accordingly.}
\label{fig:Chi2_params}
\end{figure*}


\section{Extended Dark Energy scenarios}
\label{sec:extended}

Despite the undeniable uncertainties surrounding the preliminary findings from JWST, one might wonder whether these new emerging anomalies could be somehow linked to other well-known longstanding problems in cosmology, such as the Hubble tension. This raises the question of whether they could both originate from a common issue related to our current theoretical understanding of the Universe and, on a broader scale, what kind of beyond-$\Lambda$CDM phenomenology (if any) can increase the present-day expansion rate of the Universe while also leading to a higher cumulative stellar mass density at earlier times.

At first glance, this question can even appear misplaced, as these observations are often believed to imply an older universe compared to the $\Lambda$CDM predictions. Since the age of the Universe is roughly $\propto 1/H_{0}$, this would suggest that increasing the Hubble constant could worsen the discrepancy with the observations released by JWST. However, structure formation is influenced by the evolution of primordial density perturbations and the underlying cosmology. Models addressing the Hubble tension often propose modifications at both the background level and in perturbation dynamics. This could allow for comparable or greater structure formation in a younger Universe (various N-body simulations of extension to $\Lambda$CDM show such variations, e.g.~\cite{Maio:2006zs,Baldi:2008ay,Baldi:2012ky,Barreira:2013xea,Maio:2014qwa,Adamek:2017uiq}). Furthermore, in beyond-$\Lambda$CDM models different correlations among cosmological parameters can shift their fitting values. These effects may significantly impact parameters related to structure formation, such as the matter density $\Omega_m$, and the other matter clustering parameters $\sigma_8$ and $S_8 = \sigma_8 \cdot (\Omega_m/0.3)^{1/2}$. These correlations are crucial as they could affect the amplitude and shape of the matter power spectrum.

An exercise certainly useful for understanding which kind of phenomenology could hit two targets with one arrow -- increasing $H_0$ and aligning more closely with JWST --  is  breaking down the problem into smaller parts. In particular, we focus on the baseline $\Lambda$CDM model fixing all parameters to the best-fit values provided by Planck and altering each one individually within a 4-standard-deviation range. Through this analysis, focusing on the JWST likelihood ($\chi^2_{\rm JWST}$), we identify physical adjustments for better consistency with observations. From \autoref{fig:Chi2_params} we can derive a quite significant amount of information:

\begin{itemize}
    \item First and foremost, we observe that the parameter on which $\chi^2_{\rm JWST}$ is most sensitive is the matter density parameter $\Omega_m$. In particular, a larger fraction of matter in the Universe will considerably improve the quality of the fit to JWST observations by facilitating structure formation. \footnote{For similar discussions involving quasars at high redshifts, see, e.g., Refs.~\cite{Risaliti:2018reu,Yang:2019vgk}.}
    \item Secondly, we can clearly note that increasing the amplitude of the primordial perturbations $A_s$ or considering a larger tilt $n_s$ results in a significant reduction in $\chi^2_{\rm JWST}$. This is in line with previous findings documented in Ref.~\cite{Forconi:2023izg}, where it was argued that relaxing the Planck constraints on polarization, which in turns allows $\tau$ to reach considerably higher values~\cite{Giare:2023ejv}, can substantially improve the agreement between JWST and CMB data\footnote{Since there exists a well-known degeneracy relation $A_s e^{-2\tau}$, high values of $\tau$ can be compensated by higher values of $A_s$}. Similarly, larger $n_s$ can substantially increase the power in the matter power spectrum on small scales, also facilitating more structure to form.
    \item Finally, concerning $H_0$, we observe that a straightforward increase in the value of this parameter while simultaneously keeping $\sigma_8$ constant worsens the fit to JWST data. This aligns with the argument that, when fixing structure formation parameters, a younger Universe reduces the number of structures that can form. However, the impact of $H_0$ on $\chi^2_{\rm JWST}$ is relatively small and it proves how the Hubble constant plays only a partial role in a more complex interplay of various parameters.
\end{itemize}

Summarizing these results, from a phenomenological standpoint, an effective model to increase the value of $H_0$ and improve the agreement with preliminary JWST data should predict a higher spectral index, along with a greater quantity of matter in the Universe and possibly higher values of $\sigma_8$ and $S_8$. On one side, this phenomenology is common in proposals aimed at resolving the Hubble tension by introducing new physical components that act before recombination. For this reason, we explore extensions related to the early Universe and, as a case study, analyze EDE cosmology. On the other hand, larger values of matter clustering parameters $\sigma_8$ and $S_8$ can also be achieved within late-time solutions of the Hubble tension that attempt to modify physics after recombination, influencing the value of $H_0$ derived from the angular distance from the CMB. Therefore, in the spirit of not leaving anything untried, we also test IDE cosmologies where both the growth of perturbations and the matter clustering are significantly different than in $\Lambda$CDM. Below, we summarize the basic aspects of the theoretical models we will be considering.

\subsection{Early Dark Energy}
Early Dark Energy models are a natural hypothesis of dark energy, see e.g., Refs.~\cite{Wetterich:2004pv,Doran:2006kp,Hollenstein:2009ph,Calabrese:2010uf,Calabrese:2011hg,Calabrese:2011hg,Pettorino:2013ia,Archidiacono:2014msa,Poulin:2023lkg,Poulin:2018dzj,Poulin:2018zxs,Smith:2019ihp,Niedermann:2019olb,Niedermann:2020dwg, Murgia:2020ryi,Ye:2020btb,Klypin:2020tud,Hill:2020osr,Herold:2021ksg,Herold:2022iib,Reeves:2022aoi,Jiang:2022uyg,Simon:2022adh,Smith:2022hwi,Kamionkowski:2022pkx,Niedermann:2023ssr,Cruz:2023lmn,Eskilt:2023nxm,Smith:2023oop,Sharma:2023kzr,Efstathiou:2023fbn,Gsponer:2023wpm,Goldstein:2023gnw}. Deviating from the traditional cosmological constant framework, EDE models  account for a non-negligible contribution from dark energy in the early Universe. In addition, these EDE models can be based on a generic dark energy fluids which are inhomogeneous. Their density and pressure vary over time, leading to a non-static equation of state. The phenomenological analyses
of these inhomogeneous dark energy models usually require additional dark energy clustering parameters, the dark energy effective sound speed and the dark energy anisotropic stress.  The effective sound speed determines the clustering properties of dark energy and consequently it affects the growth of matter density fluctuations. Therefore, in principle, its presence could be revealed in large scale structure observations. The growth of perturbations can also be affected by the anisotropic stress contributions  which lead to a damping in the velocity perturbations. 

Recently, EDE models have garnered significant attention, particularly due to their potential role in addressing some of the aforementioned cosmological tensions~\cite{Kamionkowski:2022pkx,Poulin:2023lkg,Abdalla:2022yfr}. Our analysis will concentrate on the EDE implementation detailed in~\cite{Hill:2020osr}. This model proposes that, in the early Universe, a light scalar field deviates from its potential minimum and, constrained by Hubble friction, is functionally similar to a cosmological constant. As soon as, at some particular redshift $z_\star$, the Hubble parameter reduces to be less than the mass of the field, the scalar field rolls down its potential and begins to oscillate about the minimum. To avoid spoiling late-time
cosmology, the vacuum energy must redshift away quicker than matter (i.e., faster than $a^{-3}$), and the field should behave as a subdominant component. A typical set of parameters used in this model is: the fractional contribution to the total energy density of the Universe, $f_{\rm EDE}(z) \equiv \rho_{\rm EDE}(z)/\rho_{\rm tot}(z)$ evaluated at the critical redshift $z_c$ at which it reaches the maximum value, and $\theta_i$, which is the parameter that usually describes the initial field displacement. This particular behavior implies a larger amount of energy-density in the early Universe (just prior to recombination), a reduction of the sound horizon and, consequently, a larger value of the Hubble constant inferred by CMB observations. This is the reason why EDE models have been proposed as a possible solution to the Hubble constant tension.

\subsection{Interacting Dark Energy}
Interacting Dark Energy models describe a phenomenological scenario where the dark fluids of the Universe interact with each other by allowing a transfer of energy and/or momentum between them, see e.g., Refs.~\cite{Valiviita:2008iv,Gavela:2009cy,Salvatelli:2014zta,Sola:2016jky,DiValentino:2017iww,Kumar:2017dnp,Wang:2016lxa,SolaPeracaula:2017esw,Sola:2017znb,Gomez-Valent:2018nib,Martinelli:2019dau,Yang:2019uog,DiValentino:2019ffd,Pan:2019jqh,Kumar:2019wfs,Yang:2018euj,Escudero:2015yka,Kumar:2016zpg,Murgia:2016ccp,Pourtsidou:2016ico,Yang:2018ubt,Barros:2018efl,Yang:2019uzo,Pan:2019gop,DiValentino:2019jae,DiValentino:2020leo,Yao:2020pji,Lucca:2020zjb,DiValentino:2020kpf,Gomez-Valent:2020mqn,Yang:2020uga,Yao:2020hkw,Pan:2020bur,DiValentino:2020vnx,Hogg:2020rdp,SolaPeracaula:2021gxi,Lucca:2021dxo,Kumar:2021eev,Yang:2021hxg,Gao:2021xnk,Yang:2021oxc,Lucca:2021eqy,Halder:2021jiv,Kaneta:2022kjj,Gariazzo:2021qtg,Nunes:2021zzi,Yang:2022csz,Nunes:2022bhn,Goh:2022gxo,Gomez-Valent:2022bku,vanderWesthuizen:2023hcl,Zhai:2023yny,Bernui:2023byc,deCruzPerez:2023wzd,Escamilla:2023shf}. Instead, the other components of the Universe (such as radiation and baryons) remain unaffected. The background evolution for dark energy and dark matter is modified, as the continuity equations for the single component present an interaction function $Q$ whose sign governs the energy-momentum flow. A negative value of the interaction rate, $Q < 0$, implies a transfer of energy and/or momentum from pressureless dark matter to dark energy, while the opposite, refers to an energy-momentum flow from the dark energy sector to the dark matter one. In order to solve the background evolution, one would need a specific interaction function $Q$. Depending on such a function, it can be solved either analytically or numerically, together with the equation for the Hubble rate evolution. In what follows we shall use the well-known interaction rate~\cite{He:2008si,Valiviita:2008iv,Gavela:2009cy,Gavela:2010tm,Honorez:2010rr}:    
\begin{eqnarray}
Q =  \xi {\cal H} \rho_{\rm de}\,,
\label{eq:coupling}
\end{eqnarray}
where $\xi$ is a dimensionless coupling parameter. The equation governing the evolution of the density perturbations for the dark sector can be found in~\cite{Valiviita:2008iv,Gavela:2009cy,Gavela:2010tm}. IDE models may suffer from instabilities in the perturbation evolution~\cite{Valiviita:2008iv,He:2008si}. Our analysis adheres to the criterion of Ref.~\cite{Gavela:2009cy} in terms of the so-called \emph{doom factor} 
\begin{equation}
\textbf{d}= \frac{Q}{3\mathcal{H}(1+w) \rho_{\rm de}}~,
\end{equation}
which is required to be negative in order to avoid instabilities. Consequently this stability condition for our case is translated into a stable parameter space in which $(1+w)$ and $\xi$ must have opposite sign~\cite{Gavela:2009cy}. Therefore, in the phantom regime in which $(1+w)$ is a negative quantity the dimensionless coupling $\xi$ must be positive. On the other hand, in the quintessence region $\xi$ must be negative. For earlier studies, see~\cite{Valiviita:2008iv,He:2008si,Jackson:2009mz,Gavela:2010tm,Clemson:2011an,Li:2014eha,Li:2014cee,Guo:2017hea,Zhang:2017ize,Guo:2018gyo,Yang:2018euj,Dai:2019vif}.

We conclude this section with a final remark: even if the interaction scenario considered here is a pure  phenomenological model, some studies have shown that using a multi-scalar field action, the coupling function (\ref{eq:coupling}) can be derived~\cite{Pan:2020zza}. Therefore, the interaction model of the form given by Eq.~(\ref{eq:coupling}) also benefits from a solid theoretical motivation. 

\section{Methodology}
\label{sec:methodology}
\subsection{Theory}
To explore the extended dark energy scenarios in relation to the JWST observations, we strictly follow the methodology detailed in Ref.~\cite{Forconi:2023izg}. We compute the predicted Cumulative Stellar Mass Density (CSMD hereafter), which is given by
\begin{equation}
    \rho_\star(M_\star)\leq\epsilon f_b\int^{z_2}_{z_1}\int^{\infty}_{M_h}\frac{dn}{dM}MdM\frac{dV}{V(z_1,z_2)}~,
    \label{eq:CSMD}
\end{equation}
where we have $M_{\star}=\epsilon f_b M_{\rm h}$ with $M_\star$ the mass of the galaxy, $M_{\rm h}$ is the halo mass, $f_b=\Omega_b/\Omega_m$ is the cosmic baryon fraction and $\epsilon$ is the star formation efficiency of converting baryons into stars. For our analysis, we opt for a conservative approach and fix $\epsilon=0.2$ following Ref.~\cite{Tacchella:2018qny}. However, as suggested by  Ref.~\cite{Tacchella:2018qny}, in principle star formation efficiency can be a function of the halo mass and further adjustments to star formation physics might be needed for more precise computations~\cite{2023arXiv231114804C}. Nonetheless, for the mass scale we are working with, $\epsilon$ can only vary smoothly as a function of mass, following a power law. Therefore, given the short mass range we are using in our analyses, this does not change in a significant way our main conclusions. Notice also that we consider the cosmic baryon fraction instead of computing the baryon evolution in different halos~\cite{Allen:2004cd,Vikhlinin:2005mp,Kravtsov:2005ab,Borgani:2009cd,Mantz:2014xba,Maio:2014qwa,Panchal:2024dcl}\footnote{ Notice that, with the baryon fraction $f_b$ playing the role of a multiplicative factor in front of the integral \eqref{eq:CSMD}, if $f_b$ increases, the theoretical predictions are proportionally pushed towards the observed data points. Just for reference, fixing $f_b$ as large as $f_b = 0.23$, we find an expected decrease in the $\chi^2$ of about 55\%.}. All these methodological choices and simplifications are widely used in the literature and allow us to present conservative and credible results that, without sacrificing generality, can be directly compared with similar works following the same approach.

Furthermore, $\frac{dn(M,z)}{dM}$ is the comoving number density of the collapsed objects between a mass range from $M$ to $M+\Delta M$ at a certain redshift, given by

\begin{equation}
    \frac{dn(M,z)}{dM}=F(\nu)\frac{\rho_m}{M^2}\left\lvert\frac{d\ln\sigma(M,z)}{d\ln M}\right\rvert~,
    \label{eq:HMassFunction}
\end{equation}
\noindent where $\rho_m$ is the comoving matter density of the Universe, $F(\nu)$ the  Sheth-Tormen (ST) Halo mass function \cite{Sheth:1999mn,Sheth:1999su} and $\sigma(R,z)$ is the variance of linear density field smoothed at a scale $R$, where we assume a Top-Hat window function. The ST formalism is theoretically motivated in terms of the collapse of halos \cite{Maggiore_2010,Achitouv_2012} and has been exhaustively tested by N-body simulations for different dark energy models taking different priors for $\Omega_m$ and $\Omega_{\Lambda}$ \cite{despali_2015}. In this study, we assume that the halo mass function of the ST type applies to both theoretical frameworks examined. Nevertheless, analyses of future JWST data may need to consider a more sophisticated fit.

\subsection{Observational data}
\label{sec:obsdata}
Regarding the JWST observations, given their preliminary nature and the wide research interest they have generated, a multitude of datasets obtained following different methodologies have been released. Consequently, a choice of which dataset to use becomes necessary. In this study, we consider four independent datasets summarized in \autoref{tab:obs_JWST} and listed below: Notice that many additional measurements have been released, see, e.g., Ref.~\cite{Bouwens_2023,xiao2023massive}. In future, these measurements could potentially serve as independent tests of our preliminary findings.

\begin{itemize}
\item Confirmed CSMD measurements taken from Ref.~\cite{2023Natu}. We find this dataset particularly well-suited for our analysis as it provides explicit constraints on the CSMD that are directly linked to cosmological structure formation. For this dataset, the errors reported in~\autoref{tab:obs_JWST} are assumed to follow a Log Normal distribution. We refer to this dataset as \textbf{\textit{JWST-CEERS}}. 

\item Five observational CSMD coming from the photometric data of JWST coverage of the UKIDSS Ultra Deep Survey (UDS) and Hubble Ultra Deep Field (HUDF)~\cite{Navarro-Carrera:2023ytd}. The errors for HUDF $\&$ UDS displayed in~\autoref{tab:obs_JWST} are conservatively taken as the maximum value, while the stellar masses are set to be equal to $M=10^8M_\odot$ as suggested in Table 6 of Ref.~\cite{Navarro-Carrera:2023ytd}. We refer to this dataset as \textbf{\textit{JWST-HUDF\,$\&$UDS}}. 

\item Two values of the observational CSMD coming from optical data in the GLASS-ERS 1324 program~\cite{Santini_GLASS_Mstar}. The stellar mass for GLASS datapoints is set to the average between the mass interval reported in~\autoref{tab:obs_JWST} where we refer to this dataset as \textbf{\textit{JWST-GLASS}}.

\item JWST FRESCO NIRCam/grism survey~\cite{xiao2023massive}. This dataset spans an area of 124 arcmin$^2$, covering a survey volume of approximately $1.2 \times 10^6$ Mpc$^3$ within the redshift range $z \in [5,9]$. We refer to it as \textbf{\textit{JWST-FRESCO}}. Taking the \textit{JWST-FRESCO} data at face value (and barring any potential systematic errors), we consider 3 obscured galaxies located within densely dusty regions, with redshifts in the range $5 \lesssim z \lesssim 6$. Referring to Fig. 3 of Ref.~\cite{xiao2023massive}, we can see that these galaxies show exceptionally extreme properties such as dark matter halo masses of $\log{\left( M_{\mathrm{halo}} / M_\odot \right)} = 12.88^{+0.11}_{-0.13}$, $12.68^{+0.23}_{-0.17}$, and $12.54^{+0.17}_{-0.18}$. Notice also that the quantity measured by \textit{JWST-FRESCO} is the cumulative comoving number density of dark matter halos, not the CSMD. For any given model of cosmology, the cumulative comoving number density of dark matter halos can be computed as:
\begin{equation}
    n(>M_{\mathrm{halo}}) =  \int_{z_1}^{z_2} \int_{M_{\rm{halo}}}^{\infty} \frac{dn}{dM} \, dM,
    \label{eq:CNDM}
\end{equation}
where the integrand in Eq.~\eqref{eq:CNDM} is defined by Eq.~\eqref{eq:HMassFunction}. Regarding the error for the FRESCO dataset, we use the approach of Poissonian approximation for small numbers of observed events (the interested reader can refer to our Appendix~\ref{app:FRESCO} for more details). 
\end{itemize}

\begin{table}[h]
    \centering
    \renewcommand{\arraystretch}{1.5}
    \begin{tabular}{|c|c|c|c|}
        \hline
        $z$ & $\ln{\left(\rho/M_\odot Mpc^{-3}\right)}$ & $\ln{\left(M/M_\odot\right)}$ & Dataset \\
        \hline
        \multirow{2}{*}{$7<z<8.5$} & $5.893\pm0.345$ & $10.030$ & \multirow{4}{*}{CEERS~\cite{2023Natu}} \\
         & $5.676\pm0.652$ & $10.75$  & \\
         \cline{1-3}
        \multirow{2}{*}{$8.5<z<10$} & $5.709\pm0.386$ & $9.704$ & \\
         & $5.386\pm0.653$ & $10.408$  & \\
        \hline
        \hline
        $3.5<z<4.5$ & $7.00^{+0.14}_{-0.16}$ &$10.48\pm0.15$  & \multirow{5}{*}{HUDF $\&$ UDS~\cite{Navarro-Carrera:2023ytd}} \\
         \cline{1-3}
         $4.5<z<5.5$ & $6.79^{+0.20}_{-0.28}$ & $10.45\pm0.27$  & \\
         \cline{1-3}
        $5.5<z<6.5$ & $6.67^{+0.21}_{-0.23}$ & $10.33\pm0.36$ & \\
        \cline{1-3}
        $6.5<z<7.5$ & $6.51^{+0.42}_{-0.60}$ & $10.68\pm0.79$  & \\
         \cline{1-3}
        $7.5<z<8.5$ & $5.75^{+0.59}_{-0.1.10}$ &[10.70] & \\
        \hline
        \hline
        $6.9<z<8.5$ & $5.07\pm0.52$ &\multirow{2}{*}{$7.2< \ln{\left(M/M_\odot\right)} <9.3$}  & \multirow{2}{*}{GLASS~\cite{Santini_GLASS_Mstar}} \\
        \cline{1-2}
        $3.5<z<4.5$ & $4.52\pm0.65$ &  &  \\
        \hline
        \hline
        & $\ln{\left(n(>M_{\rm halo}\right)}$ & $\ln{\left(M_{halo}/M_\odot\right)}$ & \\
         \cline{2-3}
        \multirow{3}{*}{$5<z<6$} & \multirow{3}{*}{$-5.52^{+0.69}_{-0.58}$} & $12.88^{+0.11}_{-0.13}$ & \multirow{3}{*}{FRESCO~\cite{xiao2023massive}} \\
        \cline{3-3}
         &  & $12.68^{+0.23}_{-0.17}$  & \\
         \cline{3-3}
         &  & $12.54^{+0.17}_{-0.18}$  & \\
         \hline
        
    \end{tabular}
    \caption{Observational points for JWST for the four different dataset. The values here must be rescaled by the corresponding comoving volume and luminosity distance for the Planck bestfit $\Lambda$CDM model.}
    \label{tab:obs_JWST}
\end{table}
We highlight that in \eq{CSMD}, $V(z_1,z_2) = \frac{4}{3} \pi \left[R^3(z_2) - R^3(z_1)\right]$ represents the model-dependent comoving volume of the Universe between redshifts $z_1$ and $z_2$ (with $R(z)$ being the comoving radius at $z$). In the values of $\rho^{\text{obs}}$ given in \autoref{tab:obs_JWST}, the comoving volume has been derived assuming the best-fit $\Lambda$CDM model (\texttt{Planck TTTEEE+lowE+lensing} CMB measurements with $h=0.6732$, $\Omega_m=0.3158$, $n_s=0.96605$, $\sigma_8=0.8120$, see~\cite{2023Natu}), making the data points model-dependent, too. As a result, these values need to be appropriately rescaled before interfacing them with the predicted values for dark energy scenarios. That is, they must be corrected by a factor equal to the ratio of the two comoving volumes ($V_{\Lambda}/V_{\rm DE}$). Similarly, the conversion between luminosity fluxes and distances for the stellar masses has been inferred, again, under the underlying assumption of the best-fit Planck $\Lambda$CDM model. A similar rescaling has to be applied, using the squared ratios of the luminosity distances.

\subsection{Numerical analyses}
\label{sec:num}
For our analyses, we perform a Monte Carlo Markov Chain (MCMC)  using the publicly available package \texttt{Cobaya}~\cite{Torrado:2020dgo} and generate theoretical predictions exploiting with a modified version of the software \texttt{CLASS}~\cite{2011arXiv1104.2932L,2011JCAP07034B} to address the IDE scenario, while, for EDE, the publicly available software \texttt{CLASS EDE},\footnote{\url{https://github.com/mwt5345/class_ede}} which solves the evolution of the background and of the perturbations in the presence of a scalar field by means of the Klein–Gordon equation. We investigate the posterior distributions of our parameter space through the MCMC sampler developed for CosmoMC~\cite{Lewis:2002ah,Lewis:2013hha} and tailored for parameter spaces with a speed hierarchy which also implements the ”fast dragging” procedure~\cite{neal2005taking}.  

The likelihood used for the MCMC analysis are:
\begin{itemize}
\item CMB temperature and polarization power spectra from the legacy Planck release~\cite{Planck:2018vyg,Planck:2019nip} with  \textit{plik} TTTEEE+low-$\ell$+lowE.
\item Lensing Planck 2018 likelihood~\cite{Planck:2018lb}, reconstructed from measurements of the power spectrum of the lensing potential.
\end{itemize}
In the following discussion, we will refer to the combinations of these two datasets simply as CMB. The convergence of the chains obtained with this procedure is assessed using the Gelman-Rubin criterion~\cite{Gelman:1992zz} setting a convergence threshold at $R-1\lesssim 0.02$.

Once the chains have converged,\footnote{The converged chains are taken with 50$\%$ of burn-in.} for each sampled model we calculate the CMSD (or the comoving number density of dark matter halos for the \textit{JWST-FRESCO} mesurements)  corresponding to different parameter combinations explored in the MCMC analysis. In particular, we set the cosmology by varying the cosmological parameters $\{\Omega_b,\Omega_c,\theta_s,\tau\}$, the inflationary parameters $\{n_s,A_s\}$ and the extra parameters from EDE $\{f_{\rm EDE},\log_{10}z_c,\theta_i\}$ and IDE $\{w,\xi\}$. Afterwards, \texttt{CLASS} is employed to compute the linear matter power spectrum $P(k,z)$ and subsequently the variance and, taking its derivative, we estimate the Halo mass function, see \eq{HMassFunction}. 
Eventually, through double integration over mass and redshift, \eq{CSMD}, we arrive at $\rho_\star(M)$. For each point in the MCMC chains obtained within a given cosmological model, we calculate the $\chi^2_{\rm JWST}$ for various datasets listed in \autoref{tab:obs_JWST}. Each dataset corresponds to a distinct $\chi^2_{\rm JWST}$. Specifically, for any given dataset, $\chi^2_{\rm JWST}$ is determined straightforwardly by
\begin{equation}
\chi^2_{\rm JWST}=\sum_i (x_i^{\rm th}-\bar{x}_i)^2/\sigma_i^2,
\label{eq:chi2_forumla}
\end{equation}
where, $\bar{x}_i$ is the observed value, $\sigma_i^2$ is its error, $x_i^{\rm th}$ the theoretically predicted quantity and $i$ runs over the number of data points of the specific JWST dataset. We then obtain the updated constraints on the cosmological parameters by re-weighting the MCMC chains, i.e. performing an importance sample, using the package \texttt{getdist}.

\section{Results}
\label{sec:results}
We now discuss the results obtained for the various extended models of Dark Energy considered in this study. To promote better organization of our findings, we divide this section into two minor subsections: in \autoref{sec.EDE} we focus on the results obtained for EDE, while in \autoref{sec.IDE} we present the findings related to IDE.

\subsection{Results for Early Dark Energy}
\label{sec.EDE}

We start by examining EDE. For this model, we summarize the best-fit values related to the most relevant parameters in \crefrange{tab:EDE}{tab:EDE_Fresco}, distinguishing the results obtained by considering the four JWST datasets independently.
\begin{table}
	\begin{center}
		\renewcommand{\arraystretch}{1.5}
		\begin{tabular}{|c|  c | c |c |c| }
  	        \hline
			\textbf{Parameter} &\textbf{CMB}  &  & \textbf{JWST-CEERS}  & \textbf{CMB+JWST}\\
        \hline
             \multirow{3}{*}[+2ex]{$n_s$}&\multirow{3}{*}[+2ex]{$0.981$}&$z_{\rm low}$&$0.997$&$0.981$\\
			 &&$z_{\rm high}$&$0.997$&$0.981$\\
        \hline
             \multirow{3}{*}[+2ex]{$H_0$}&\multirow{3}{*}[+2ex]{$69.45$}&$z_{\rm low}$&$72.60$&$69.45$\\
			 &&$z_{\rm high}$&$72.60$&$69.45$\\
        \hline
             \multirow{3}{*}[+2ex]{$\sigma_8$}&\multirow{3}{*}[+2ex]{$0.8273$}&$z_{\rm low}$&$0.854$&$0.8273$\\
			 &&$z_{\rm high}$&$0.854$&$0.8273$\\
        \hline
             \multirow{3}{*}[+2ex]{$\tau$}&\multirow{3}{*}[+2ex]{$0.0575$}&$z_{\rm low}$&$0.0497$&$0.05753$\\
			 &&$z_{\rm high}$&$0.0497$&$0.05753$\\
        \hline
             \multirow{3}{*}[+2ex]{$\Omega_m$}&\multirow{3}{*}[+2ex]{$0.307$}&$z_{\rm low}$&$0.304$&$0.307$\\
			 &&$z_{\rm high}$&$0.304$&$0.307$\\
        \hline
             \multirow{3}{*}[+2ex]{$f_{\rm EDE}$}&\multirow{3}{*}[+2ex]{$0.0628$}&$z_{\rm low}$&$0.151$&$0.0628$\\
			 &&$z_{\rm high}$&$0.151$&$0.0628$\\
        \hline
            \multirow{3}{*}[+2ex]{$\chi^2$}&\multirow{3}{*}[+2ex]{$2772$}&$z_{\rm low}$&$5.75$&$2782.76$ $(2772+10.76)$\\
			 &&$z_{\rm high}$&$7.99$&$2787.34$ $(2772+15.34)$\\
        \hline
\end{tabular}
	\end{center}
	\caption{Results for EDE. We provide the best-fit values of cosmological parameters, namely the combination that minimizes the $\chi^2$ of the fit to the CMB data alone ($\chi^2_{\rm CMB}$), \textit{JWST-CEERS} data alone ($\chi^2_{\rm JWST-CEERS}$), and CMB+JWST-CEERS data ($\chi^2_{\rm CMB+JWST}$).}
	\label{tab:EDE}
\end{table}
\begin{table}
	\begin{center}
		\renewcommand{\arraystretch}{1.5}
		\begin{tabular}{|c|  c  |c |c| }
  	        \hline
			\textbf{Parameter} &\textbf{CMB}  & \textbf{JWST-HUDF$\&$UDS}  & \textbf{CMB+JWST}\\
        \hline
             $n_s$&$0.981$&$0.997$&$0.997$\\
        \hline
             $H_0$&$69.45$&$73.42$&$73.46$\\
        \hline
             $\sigma_8$&$0.8273$&$0.854$&$0.8498$\\
        \hline
            $\tau$&$0.0575$&$0.0497$&$0.05179$\\
        \hline
             $\Omega_m$&$0.307$&$0.295$&$0.2947$\\
        \hline
             $f_{\rm EDE}$&$0.0628$&$0.160$&$0.159$\\
        \hline
            $\chi^2$&$2772$&$46$&$2829$ $(2783+46)$\\
        \hline
\end{tabular}
	\end{center}
	\caption{Results for EDE. We provide the best-fit values of cosmological parameters, namely the combination that minimizes the $\chi^2$ of the fit to the CMB data alone ($\chi^2_{\rm CMB}$), \textit{JWST-HUDF$\&$UDS}~\cite{Navarro-Carrera:2023ytd} data alone ($\chi^2_{\rm JWST-HUDFUDS}$), and CMB+\textit{JWST-HUDF$\&$UDS} data ($\chi^2_{\rm CMB+JWST}$).}
	\label{tab:EDE_HUDF}
\end{table}
\begin{table}
	\begin{center}
		\renewcommand{\arraystretch}{1.5}
		\begin{tabular}{|c|  c  |c |c| }
  	        \hline
			\textbf{Parameter} &\textbf{CMB}  & \textbf{JWST-GLASS}  & \textbf{CMB+JWST}\\
        \hline
             $n_s$&$0.981$&$0.977$&$0.978$\\
        \hline
             $H_0$&$69.45$&$71.30$&$69.88$\\
        \hline
             $\sigma_8$&$0.8273$&$0.8274$&$0.8267$\\
        \hline
            $\tau$&$0.0575$&$0.03343$&$0.05103$\\
        \hline
             $\Omega_m$&$0.307$&$0.312$&$0.312$\\
        \hline
             $f_{\rm EDE}$&$0.0628$&$0.130$&$0.0853$\\
        \hline
            $\chi^2$&$2772$&$20$&$2797$ $(2774+23)$\\
        \hline
\end{tabular}
	\end{center}
	\caption{Results for EDE. We provide the best-fit values of cosmological parameters, namely the combination that minimizes the $\chi^2$ of the fit to the CMB data alone ($\chi^2_{\rm CMB}$), \textit{JWST-GLASS} data alone ($\chi^2_{\rm JWST-GLASS}$), and CMB+\textit{JWST-GLASS} data ($\chi^2_{\rm CMB+JWST}$).}
	\label{tab:EDE_GLASS}
\end{table}
\begin{table}
	\begin{center}
		\renewcommand{\arraystretch}{1.5}
		\begin{tabular}{|c|  c  |c |c| }
  	        \hline
			\textbf{Parameter} &\textbf{CMB}  & \textbf{JWST-FRESCO}  & \textbf{CMB+JWST}\\
        \hline
             $n_s$&$0.981$&$0.997$&$0.978$\\
        \hline
             $H_0$&$69.45$&$72.60$&$69.88$\\
        \hline
             $\sigma_8$&$0.8273$&$0.854$&$0.8267$\\
        \hline
            $\tau$&$0.0575$&$0.0497$&$0.0510$\\
        \hline
             $\Omega_m$&$0.307$&$0.304$&$0.312$\\
        \hline
             $f_{\rm EDE}$&$0.0628$&$0.151$&$0.0853$\\
        \hline
            $\chi_{\rm FRESCO}^2$&$2772$&$21.42$&$2804$ $(2774+30)$\\
        \hline
\end{tabular}
	\end{center}
	\caption{Results for EDE. We provide the best-fit values of cosmological parameters, namely the combination that minimizes the $\chi^2$ of the fit to the CMB data alone ($\chi^2_{\rm CMB}$), \textit{JWST-FRESCO} data alone ($\chi^2_{\rm JWST-FRESCO}$), and CMB+\textit{JWST-FRESCO} data ($\chi^2_{\rm CMB+JWST}$).}
	\label{tab:EDE_Fresco}
\end{table}
\begin{figure*}[htp]
	\centering
	\includegraphics[width=0.95\textwidth]{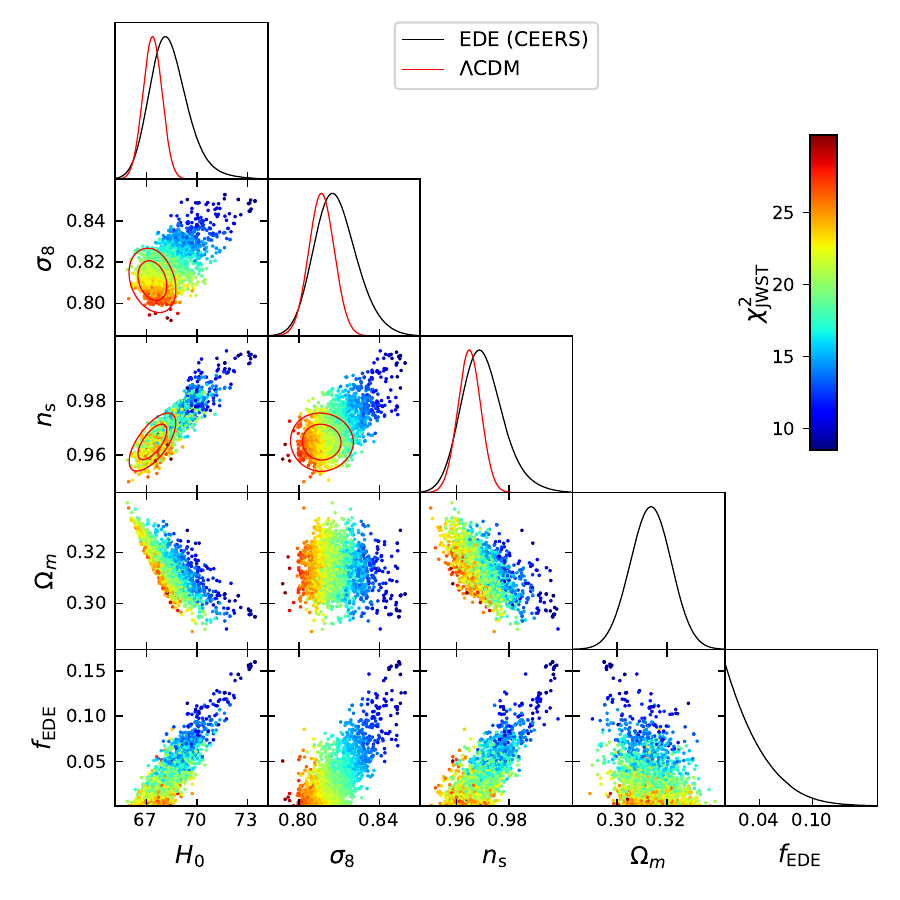}
	\caption{Triangular plot showing the distribution of points and the correlations among the most relevant parameters of EDE. The color map refers to the value of $\chi^2_{\rm JWST}$ so that the color pattern in the figure represents the direction towards which one needs to move in the parameter space to improve the fit to JWST data.}
	\label{fig:EDE_LCDM}
\end{figure*}
\begin{figure*}[htp]
	\centering
	\includegraphics[width=0.95\textwidth]{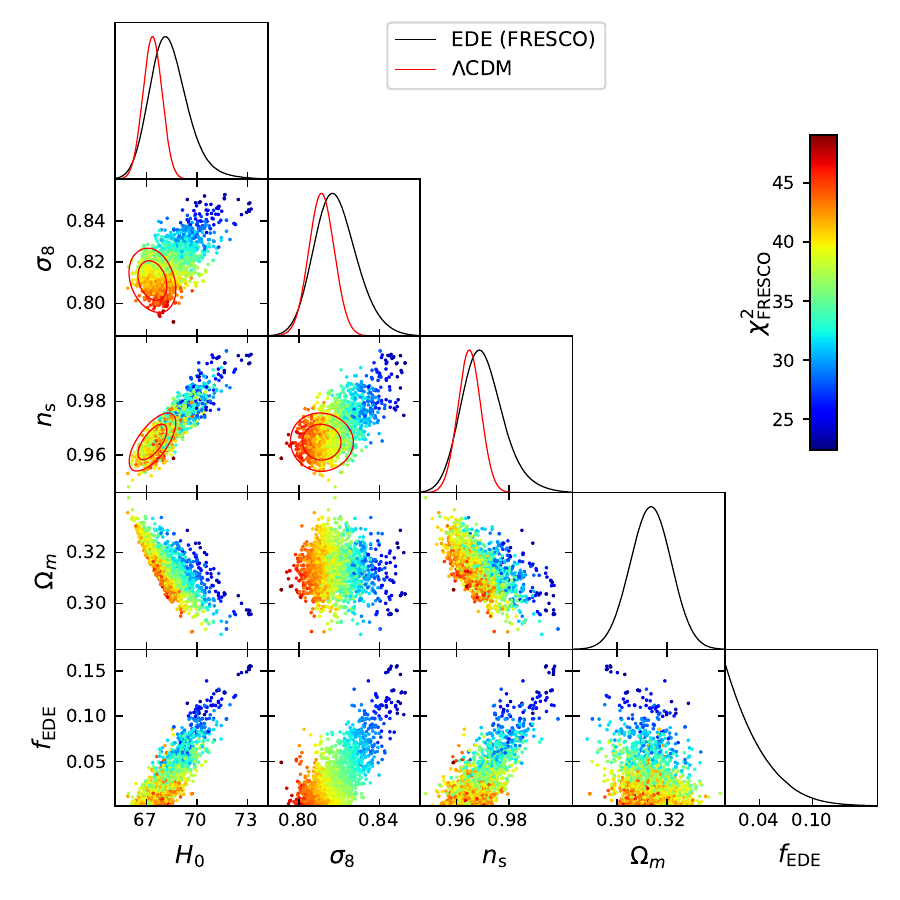}
	\caption{Triangular plot showing the distribution of points and the correlations among the most relevant parameters of EDE. The color map refers to the value of $\chi^2_{\rm FRESCO}$ so that the color pattern in the figure represents the direction towards which one needs to move in the parameter space to improve the fit to JWST FRESCO data.}
	\label{fig:EDE_LCDM_FRESCO}
\end{figure*}

First and foremost, we examine the best-fit values obtained by exclusively considering CMB measurements from the Planck satellite (indicated as CMB in the table). In this case, the values reported in \crefrange{tab:EDE}{tab:EDE_Fresco}, simply represent the combination of cosmological parameters for which $\chi^2_{\text{CMB}}$ ($=2772$) acquires its minimum value among those obtained within the MCMC analysis. It is worth noting that we retrieve results widely documented in the literature. In particular, the Planck data, while not showing any substantial preference for a non-vanishing fraction of EDE, produce a best-fit value of $f_{\text{EDE}}=0.06$. This leads to a present-day expansion rate of the Universe $H_0=69.45$ km/s/Mpc, which is generally higher than the best-fit value obtained for this parameter within the standard cosmological model. Another point that is worth emphasizing is that for the inflationary spectral index we get a best-fit value $n_s=0.981$, confirming once more the trend of EDE models in predicting a spectrum of primordial perturbations closer to the scale-invariant case than what is observed in $\Lambda$CDM and predicted by the most typical inflationary potentials~\cite{DiValentino:2018zjj,Ye:2021nej,Ye:2022efx,Jiang:2022uyg, Jiang:2022qlj,Takahashi:2021bti,Jiang:2023bsz,Peng:2023bik}
.\footnote{For other discussions surrounding the value of this parameter and the agreement among the results of different CMB probes, see, e.g., Refs.~\cite{Forconi:2021que,ACT:2020gnv,Handley:2020hdp,DiValentino:2022rdg,Giare:2022rvg,Giare:2023wzl,DiValentino:2022oon,Calderon:2023obf,Giare:2023xoc}.}

As a second step, following the methodology outlined in the previous section, for each combination of parameters in the MCMC chains (i.e., for each collected model), we calculate the $\chi^2$ against the four different JWST datasets listed in \autoref{tab:obs_JWST}. 

We start discussing the results obtained re-weighting the chains in light of $\chi^2_{\rm JWST-CEERS}$ resulting from the \textit{JWST-CEERS}  dataset. This dataset has been recently analyzed in many similar studies and allows us for direct comparison with the existing findings in the literature~\cite{Boylan-Kolchin:2022kae,Menci:2022wia}. In this case, we summarize the results in \autoref{tab:EDE}, distinguishing between the low ($z_{\rm low}$) and high ($z_{\rm high}$) redshift bins (also see \autoref{tab:obs_JWST}). Similar to the CMB analysis,  we present the specific combination of parameters that minimizes $\chi^2_{\rm JWST-CEERS}$. Furthermore, in \autoref{fig:EDE_LCDM} we provide a triangular plot showing the distribution of sampled models and the correlations among different parameters, together with a color-map representing the value of $\chi^2_{\rm{JWST-CEERS}}$. For the sake of comparison, in the same figure, we also depict the predictions of $\Lambda$CDM. A few intriguing conclusions can be derived from both \autoref{tab:EDE} and \autoref{fig:EDE_LCDM}. Firstly, there are no significant differences between the results obtained for the high and low redshift bins. Secondly, as for the best-fit values of cosmological parameters, we now find a pronounced preference for a non-vanishing fraction of EDE, $f_{\rm EDE}=0.151$. We also get higher $\sigma_8=0.85$ and observe the same trend towards higher values of inflationary spectral index $n_s=0.997$, now essentially consistent with a Harrison-Zel’dovich spectrum. As pointed out in \autoref{sec:extended}, this is exactly the kind of phenomenology one needs to increase the agreement with JWST data. Therefore -- not surprisingly -- the minimum value of $\chi^2_{\rm JWST-CEERS}$ for both the high ($\chi^2_{\rm{JWST-CEERS}}=7.99$) and low ($\chi^2_{\rm{JWST-CEERS}}=5.75$) redshift bins are significantly better than what we get in $\Lambda$CDM (where $\chi_{\rm JWST-CEERS}^2\sim 17$~\cite{Forconi:2023izg}). This suggests that EDE stands as a valid phenomenological alternative to explain (at least partially) the preliminary measurements released by \textit{JWST-CEERS}. Furthermore, regarding $H_0$, the \textit{JWST-CEERS} best-fit value reads $H_0=72.60$ km/s/Mpc. Therefore, not only within the context of EDE we can improve the agreement between the theoretical predictions of the model and the \textit{JWST-CEERS} data, but to achieve this, we move through the parameter space in the same direction needed to solve the Hubble tension, as well. This is also clearly confirmed by the color pattern in \autoref{fig:EDE_LCDM}, underscoring that it is indeed possible to address both issues within the same theoretical framework.

Finally, always in \autoref{tab:EDE}, we present the results inferred by summing up the $\chi^2$ values of CMB and \textit{JWST-CEERS}. We observe that the combination of parameters that minimizes the total $\chi^2_{\text{CMB+JWST}}$ is the same as that minimizing the fit to only the Planck data $\chi^2_{\text{CMB}}$. At first glance, this implies that the cost of improving the fit to the \textit{JWST-CEERS} data is an overall deterioration in the fit to the CMB. On the other hand, such deterioration is entirely expected, given the strong preference of Planck data for a $\Lambda$CDM cosmology and the general disagreement between \textit{JWST-CEERS} data and the latter. As extensively documented in the literature and confirmed by our analysis, individual Planck data do not provide clear evidence in favor of an EDE cosmology. In any case, the best-fit parameters suggest an inclination towards a model where the fraction of EDE remains modest and well below $f_{\rm EDE}\lesssim 0.1$. In contrast, reconciling \textit{JWST-CEERS} with an EDE cosmology would require an EDE fraction $f_{\rm EDE}\gtrsim 0.1$. Forcing such an EDE fraction into the model would source significant effects in the CMB spectra that can only be partially compensated by the observed shift in the fitting values of other cosmological parameters. Just to provide an illustrative example, the increase in the expansion rate of the Universe before recombination due to a substantial EDE component leads to a significant reduction in the value of the sound horizon at the combination, forcing the value of $H_0$ in the direction of SH0ES, which is certainly not the direction favored by CMB data. Additionally, since EDE does not alter the physics of post-recombination, a higher $H_0$ implies a lower angular diameter distance from the CMB, $D_A$. In turn, this leads to a shift in the wavenumber associated with the damping tail $k_D$, as these two parameters are related by the relationship $\ell_D \sim k_D D_A$, where the multipole $\ell_D$ is also fixed by CMB measurements. In an attempt to maintain a good fit to the damping scale, the value of $n_s$ is shifted towards a scale-invariant primordial spectrum (i.e., $n_s \to 1$) which certainly improves the agreement between \textit{JWST-CEERS} and EDE but is again highly disfavored by Planck (by over $8\sigma$ in $\Lambda$CDM). Overall, all these effects and shifts in the fitting values of cosmological parameters seem to favor \textit{JWST-CEERS} observations. However, although they partially compensate for each other, they still remain somewhat disfavored based solely on CMB data, leading to a deterioration in the fit. 

When analyzing the other JWST datasets listed in \autoref{tab:obs_JWST}, all the conclusions we have drawn so far remain mostly true. For instance, by comparing the results obtained for \textit{JWST-CEERS} in \autoref{fig:EDE_LCDM} with those obtained for \textit{JWST-FRESCO} in \autoref{fig:EDE_LCDM_FRESCO}, at first glance, we can spot the very same color patterns, indicating that a non-vanishing fraction of EDE could, in principle, help to reduce $\chi^2_{\rm FRESCO}$ while also yielding higher values of $H_0$. However, paying closer attention to the colorbar scale, we observe that as we approach the limit $f_{\rm EDE}\to 0$ moving towards the $\Lambda$CDM cosmology, we get $\chi^2_{\rm JWST-FRESCO}\sim 50$ for 3 data points. This value can be reduced all the way down to $\min(\chi^2_{\rm JWST-FRESCO})\sim 21.42$ when $f_{\rm EDE}\sim 0.15$ and $H_0\sim73$ km/s/Mpc, as seen in \autoref{tab:EDE_Fresco}. On one hand, this lends weight to the idea that EDE could potentially pave the way to partially explaining more massive galaxies and higher values of $H_0$. On the other hand, it is important to note a nearly threefold increase in $\chi^2_{\mathrm{JWST-FRESCO}}$ compared to the results obtained for \textit{JWST-CEERS}. Taking the large $\chi^2$ at face value, we must draw the conclusion that the \textit{JWST-FRESCO} dataset remains in strong disagreement with the theoretical predictions of the standard cosmological model and -- to a lesser extent -- with EDE as well. 

Looking at the $\chi^2$ of \textit{JWST-GLASS} and \textit{JWST-{HUD$\&$UDS}}, a similar conclusion can be drawn. As a result, we conclude that while EDE represents a phenomenological possibility to partially address the JWST data, it falls short of being exhaustive in fully addressing issues, leaving the quest for a more comprehensive solution wide open.

Having that said, it is worth keeping in mind some caveats surrounding the joint analyses. For instance, the total $\chi^2_{\text{CMB+JWST}}$ is obtained by considering the sum of $\chi^2$ for each sampled model in the MCMC chains \textit{afterward} and not through a joint analysis of the two experiments from the outset. Additionally, only CMB data are taken into account in the MCMC analysis, which we know do not favor high values of $f_{\text{EDE}}$ and $H_0$. Considering other datasets, such as the measurements of the expansion rate provided by the SH0ES collaboration, could lead to significantly different results in terms of the $\chi^2$ analysis, as typically pointed out by the EDE community (see, e.g., the discussion on page 25 of Ref.~\cite{Poulin:2023lkg}). Hence, a full joint likelihood analysis of all these datasets (which is beyond the aim of this work) is needed before deriving any definitive conclusions on this matter.

\subsection{Results for Interacting Dark Energy}
\label{sec.IDE}

We now move to the study of IDE. In this case, we consider three different models: the usual IDE cosmology with a fixed dark energy equation of state $w \simeq -1$, and $w$IDE models where the equation of state parameter $w$ is free to vary, although limited either in quintessence ($w > -1$) or phantom ($w < -1$) regime. For the sake of simplicity, in this subsection, we present \textit{only} the results obtained from \textit{JWST-CEERS}. Several well-motivated reasons underpin this decision. Firstly, no significant differences emerged in the results for EDE when analyzing the four different JWST datasets listed in \autoref{tab:obs_JWST}. Overall, all these observations converge on anomalous galaxies that are more massive than predicted by the standard cosmological model. Therefore, no significant disparities are anticipated when analyzing the same four datasets across the various IDE models proposed in this section. Yet another motivation involves noting that addressing these JWST anomalous observations requires a somewhat clear beyond-$\Lambda$CDM phenomenology that none of the three IDE models proposed here can offer. To streamline the analysis and emphasize this point, we focus on \textit{JWST-CEERS}, which will provide the phenomenological guidelines applicable directly to all datasets not explicitly mentioned, without exception. 

\begin{table}
	\begin{center}
		\renewcommand{\arraystretch}{1.5}
		\begin{tabular}{|c|  c | c |c |c| }
  	        \hline
			\textbf{Parameter} &\textbf{CMB}  &  & \textbf{JWST-CEERS}  & \textbf{CMB+JWST}\\
        \hline
             \multirow{3}{*}[+2ex]{$n_s$}&\multirow{3}{*}[+2ex]{$0.971$}&$z_{\rm low}$&$0.968$&$0.963$\\
			 &&$z_{\rm high}$&$0.968$&$0.961$\\
        \hline
             \multirow{3}{*}[+2ex]{$H_0$}&\multirow{3}{*}[+2ex]{$70.98$}&$z_{\rm low}$&$67.43$&$68.27$\\
			 &&$z_{\rm high}$&$67.43$&$66.69$\\
        \hline
             \multirow{3}{*}[+2ex]{$\sigma_8$}&\multirow{3}{*}[+2ex]{$1.09$}&$z_{\rm low}$&$0.878$&$0.89$\\
			 &&$z_{\rm high}$&$0.878$&$0.85$\\
        \hline
             \multirow{3}{*}[+2ex]{$\tau$}&\multirow{3}{*}[+2ex]{$0.057$}&$z_{\rm low}$&$0.0609$&$0.057$\\
			 &&$z_{\rm high}$&$0.0609$&$0.051$\\
        \hline
             \multirow{3}{*}[+2ex]{$\Omega_m$}&\multirow{3}{*}[+2ex]{$0.214$}&$z_{\rm low}$&$0.301$&$0.315$\\
			 &&$z_{\rm high}$&0.301&0.285\\
        \hline
             \multirow{3}{*}[+2ex]{$\xi$}&\multirow{3}{*}[+2ex]{$-0.28$}&$z_{\rm low}$&$-0.0734$&$-0.098$\\
			 &&$z_{\rm high}$&$-0.073$&$-0.053$\\
        \hline
             \multirow{3}{*}[+2ex]{$\chi^2$}&\multirow{3}{*}[+2ex]{$2781$}&$z_{\rm low}$&$12.50$&$2800.79$ $(2785+15.79)$\\
			 &&$z_{\rm high}$&$18.22$&$2809.29$ $(2788+21.29)$\\
        \hline
\end{tabular}
	\end{center}
	\caption{Results for IDE, with the dark energy equation of state fixed to $w\simeq -1$. We provide the best-fit values of cosmological parameters, namely the combination that minimizes the $\chi^2$ of the fit to the CMB data alone ($\chi^2_{\rm CMB}$), \textit{JWST-CEERS} data alone ($\chi^2_{\rm JWST-CEERS}$), and CMB+JWST data ($\chi^2_{\rm CMB+JWST}$).}
 \label{tab:IDE}
\end{table}
\begin{figure*}[htp]
	\centering
	\includegraphics[width=0.95\textwidth]{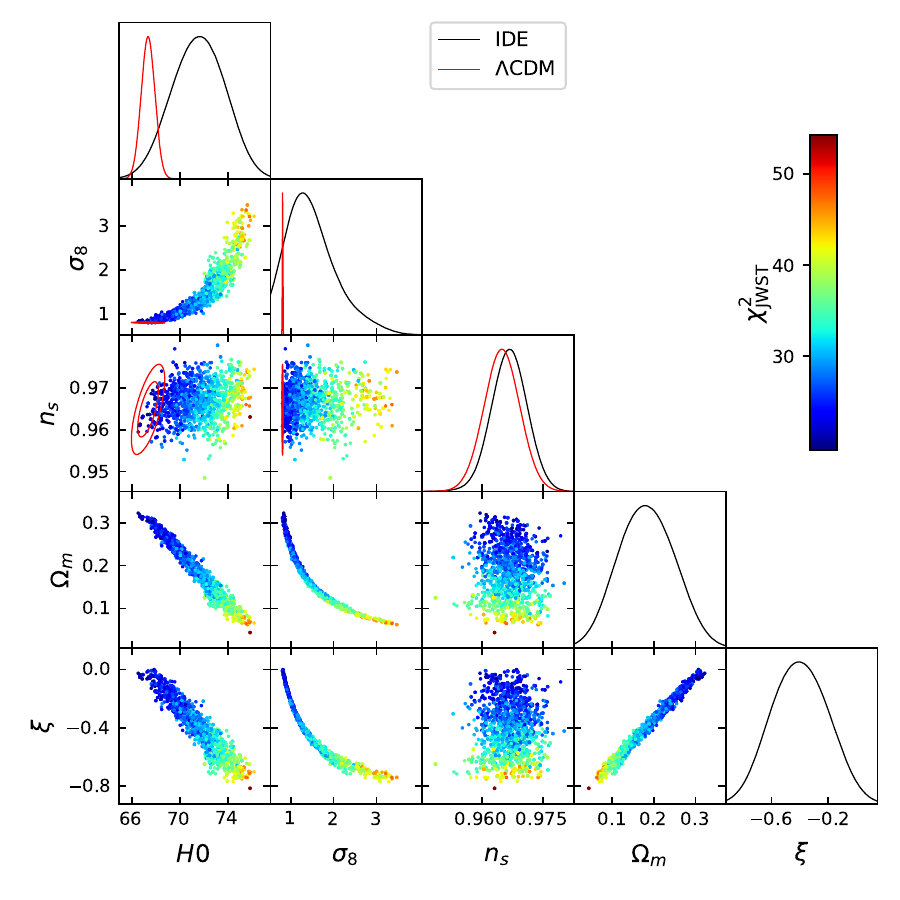}
	\caption{Triangular plot showing the distribution of points and the correlations among the most relevant parameters of IDE, when $w$ is fixed to $w\simeq -1$. The color map refers to the value of $\chi^2_{\rm JWST}$ so that the color pattern in the figure represents the direction towards which one needs to move in the parameter space to improve the fit to JWST data.}
	\label{fig:IDE_LCDM}
\end{figure*}

\autoref{tab:IDE} displays the results for the IDE model with a fixed dark energy equation of state. Similar to EDE, we consider three different combinations of data: CMB, \textit{JWST-CEERS}, and CMB+\textit{JWST-CEERS}. In the table, we always show the combination of parameters that minimizes the $\chi^2$ for these three datasets. When focusing solely on the Planck CMB data, we note that the best-fit value for the parameter encapsulating new physics -- i.e., the coupling $\xi$ -- reads $\xi=-0.28$. This suggests a quite significant transfer of energy-momentum from the dark matter sector to the dark energy sector. As widely documented in the literature, such a transfer of energy-momentum leads to a higher present-day expansion rate of the Universe, whose best-fit value reads $H_0=70.98$ km/s/Mpc (significantly higher than in the standard cosmological model). Furthermore, while we do not observe significant differences in the value of the spectral index, we notice a tendency toward higher values of $\sigma_8=1.09$. This makes the model potentially interesting for JWST. Nevertheless, the results we obtain from \textit{JWST-CEERS} data seem to indicate precisely the opposite. In contrast to the CMB fit (which  prefers $\xi < 0$), when considering only the \textit{JWST-CEERS} likelihood, the coupling parameter $\xi$ tends towards $\xi \to 0$. Consequently, we lose any ability to increase the present-day expansion rate, getting a best fit value $H_0 = 67.43$ km/s/Mpc. \autoref{fig:IDE_LCDM} further reinforces our conclusions: it is not $n_s$ but $\Omega_m$ that now assumes a critical role. When moving in the direction $\xi < 0$, the matter density undergoes a significant decrease due to the energy transfer from dark matter to dark energy. This leads to a substantial increase in the value of $\sigma_8$ to compensate for the reduced $\Omega_m$. However, as illustrated by the color pattern in \autoref{fig:IDE_LCDM}, in order to minimize $\chi^2_{\rm JWST-CEERS}$, it becomes necessary to revert to the $\Lambda$CDM framework by preventing such energy transfer, essentially moving towards $\xi\to0$. This behaviour is further supported by comparing the best-fit values of $\Omega_m$ ($\sigma_8$) for CMB and JWST: while in the former case $\Omega_m = 0.214$ ($\sigma_8=1.09$), for \textit{JWST-CEERS}, we get back to more typical values $\Omega_m = 0.301$ ($\sigma_8=0.878$). Consequently, this model fails to provide a satisfactory fit to the \textit{JWST-CEERS} observations.

\begin{table}[htbp!]
	\begin{center}
		\renewcommand{\arraystretch}{1.5}
		\begin{tabular}{|c|  c | c |c |c| }
  	        \hline
			\textbf{Parameter} &\textbf{CMB}  &  & \textbf{JWST-CEERS}  & \textbf{CMB+JWST}\\
        \hline
             \multirow{3}{*}[+2ex]{$n_s$}&\multirow{3}{*}[+2ex]{$0.9615$}&$z_{\rm low}$&$0.967$&$0.9614$\\
			 &&$z_{\rm high}$&$0.967$&$0.9614$\\
        \hline
             \multirow{3}{*}[+2ex]{$H_0$}&\multirow{3}{*}[+2ex]{$67.35$}&$z_{\rm low}$&$65.44$&$64.92$\\
			 &&$z_{\rm high}$&$65.44$&$64.92$\\
        \hline
             \multirow{3}{*}[+2ex]{$\sigma_8$}&\multirow{3}{*}[+2ex]{$0.9342$}&$z_{\rm low}$&$0.839$&$0.871$\\
			 &&$z_{\rm high}$&$0.839$&$0.871$\\
        \hline
             \multirow{3}{*}[+2ex]{$\tau$}&\multirow{3}{*}[+2ex]{$0.05742$}&$z_{\rm low}$&$0.0686$&$0.055$\\
			 &&$z_{\rm high}$&$0.0686$&$0.055$\\
        \hline
             \multirow{3}{*}[+2ex]{$\Omega_m$}&\multirow{3}{*}[+2ex]{$0.274$}&$z_{\rm low}$&$0.326$&$0.314$\\
			 &&$z_{\rm high}$&$0.326$&$0.314$\\
        \hline
             \multirow{3}{*}[+2ex]{$\xi$}&\multirow{3}{*}[+2ex]{$-0.1743$}&$z_{\rm low}$&$-0.045$&$-0.115$\\
			 &&$z_{\rm high}$&$-0.045$&$-0.115$\\
        \hline
            \multirow{3}{*}[+2ex]{$w$}&\multirow{3}{*}[+2ex]{$-0.9483$}&$z_{\rm low}$&$-0.923$&$-0.90$\\
			 &&$z_{\rm high}$&$-0.923$&$-0.90$\\
        \hline
             \multirow{3}{*}[+2ex]{$\chi^2$}&\multirow{3}{*}[+2ex]{$2774$}&$z_{\rm low}$&$11.94$&$2790.89$ $(2776+14.89)$\\
			 &&$z_{\rm high}$&$17.5$&$2797.63$ $(2778+21.63)$\\
        \hline
\end{tabular}
	\end{center}
	\caption{Results for $w$IDE, with $w>-1$ free to vary in the quintessence regime. We provide the best-fit values of cosmological parameters, namely the combination that minimizes the $\chi^2$ of the fit to the CMB data alone ($\chi^2_{\rm CMB}$), \textit{JWST-CEERS} data alone ($\chi^2_{\rm JWST-CEERS}$), and CMB+JWST data ($\chi^2_{\rm CMB+JWST}$).}
	\label{tab:nwIDE}
\end{table}

\begin{figure*}[htbp!]
	\centering
	\includegraphics[width=0.95\textwidth]{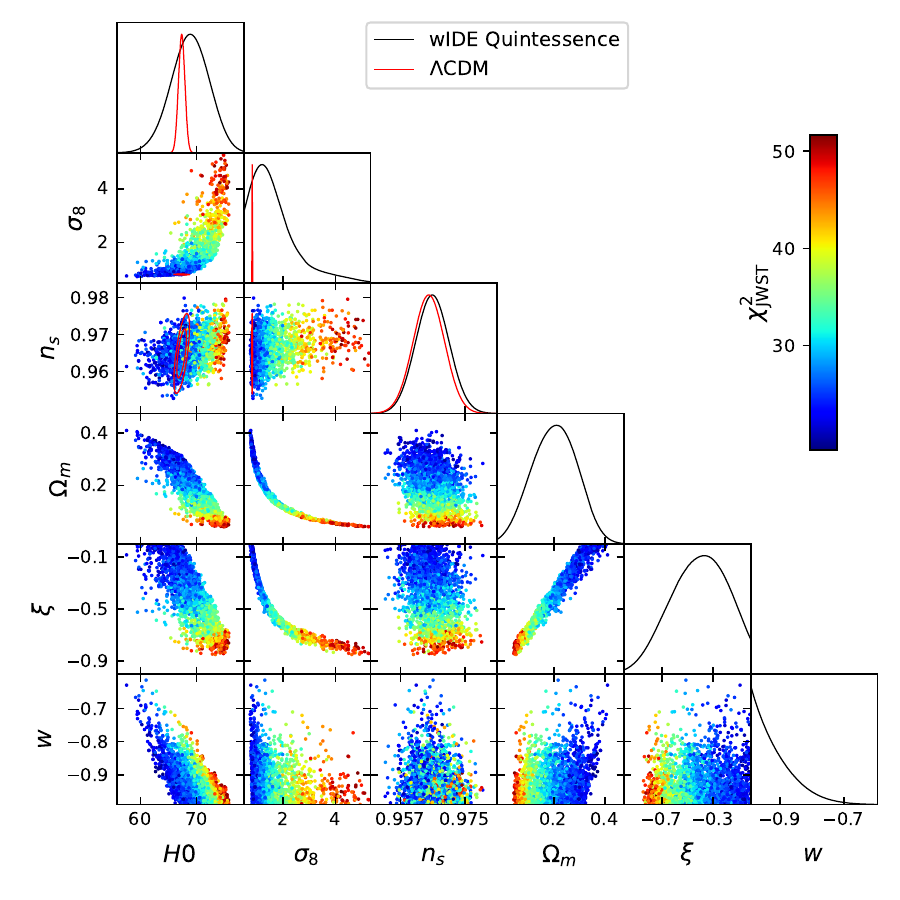}
	\caption{Triangular plot showing the distribution of points and the correlations among the most relevant parameters of $w$IDE, when $w$ is free to vary in the quintessence region $w> -1$. The color map refers to the value of $\chi^2_{\rm JWST}$ so that the color pattern in the figure represents the direction towards which one needs to move in the parameter space to improve the fit to JWST data.}
	\label{fig:nwIDE_LCDM}
\end{figure*}

\begin{table}[htbp!]
	\begin{center}
		\renewcommand{\arraystretch}{1.5}
		\begin{tabular}{|c|  c | c |c |c| }
  	        \hline
			\textbf{Parameter} &\textbf{CMB}  &  & \textbf{JWST-CEERS}  & \textbf{CMB+JWST}\\
        \hline
             \multirow{3}{*}[+2ex]{$n_s$}&\multirow{3}{*}[+2ex]{$0.9663$}&$z_{\rm low}$&$0.963$&$0.969$\\
			 &&$z_{\rm high}$&$0.967$&$0.969$\\
        \hline
             \multirow{3}{*}[+2ex]{$H_0$}&\multirow{3}{*}[+2ex]{$103.8$}&$z_{\rm low}$&$67.10$&$75.62$\\
			 &&$z_{\rm high}$&$73.88$&$75.62$\\
        \hline
             \multirow{3}{*}[+2ex]{$\sigma_8$}&\multirow{3}{*}[+2ex]{$1.026$}&$z_{\rm low}$&$0.685$&$0.760$\\
			 &&$z_{\rm high}$&$0.739$&$0.760$\\
        \hline
             \multirow{3}{*}[+2ex]{$\tau$}&\multirow{3}{*}[+2ex]{$0.05087$}&$z_{\rm low}$&$0.0612$&$0.0617$\\
			 &&$z_{\rm high}$&$0.0644$&$0.0617$\\
        \hline
             \multirow{3}{*}[+2ex]{$\Omega_m$}&\multirow{3}{*}[+2ex]{$0.139$}&$z_{\rm low}$&$0.396$&$0.292$\\
			 &&$z_{\rm high}$&$0.319$&$0.292$\\
        \hline
             \multirow{3}{*}[+2ex]{$\xi$}&\multirow{3}{*}[+2ex]{$0.05235$}&$z_{\rm low}$&$0.374$&$0.229$\\
			 &&$z_{\rm high}$&$0.299$&$0.229$\\
        \hline
            \multirow{3}{*}[+2ex]{$w$}&\multirow{3}{*}[+2ex]{$-2.0436$}&$z_{\rm low}$&$-1.149$&$-1.33$\\
			 &&$z_{\rm high}$&$-1.33$&$-1.33$\\
        \hline
             \multirow{3}{*}[+2ex]{$\chi^2$}&\multirow{3}{*}[+2ex]{$2767$}&$z_{\rm low}$&$10.8$&$2783.90$ $(2771+12.90)$\\
			 &&$z_{\rm high}$&$16.3$&$2790.15$ $(2771+19.15)$\\
        \hline
\end{tabular}
	\end{center}
	\caption{Results for $w$IDE with $w<-1$ free to vary in the phantom regime. We provide the best-fit values of cosmological parameters, namely the combination that minimizes the $\chi^2$ of the fit to the CMB data alone ($\chi^2_{\rm CMB}$), \textit{JWST-CEERS} data alone ($\chi^2_{\rm JWST-CEERS}$), and CMB+JWST data ($\chi^2_{\rm CMB+JWST}$).}
	\label{tab:pwIDE}
\end{table}

\begin{figure*}[htbp!]
	\centering
	\includegraphics[width=0.95\textwidth]{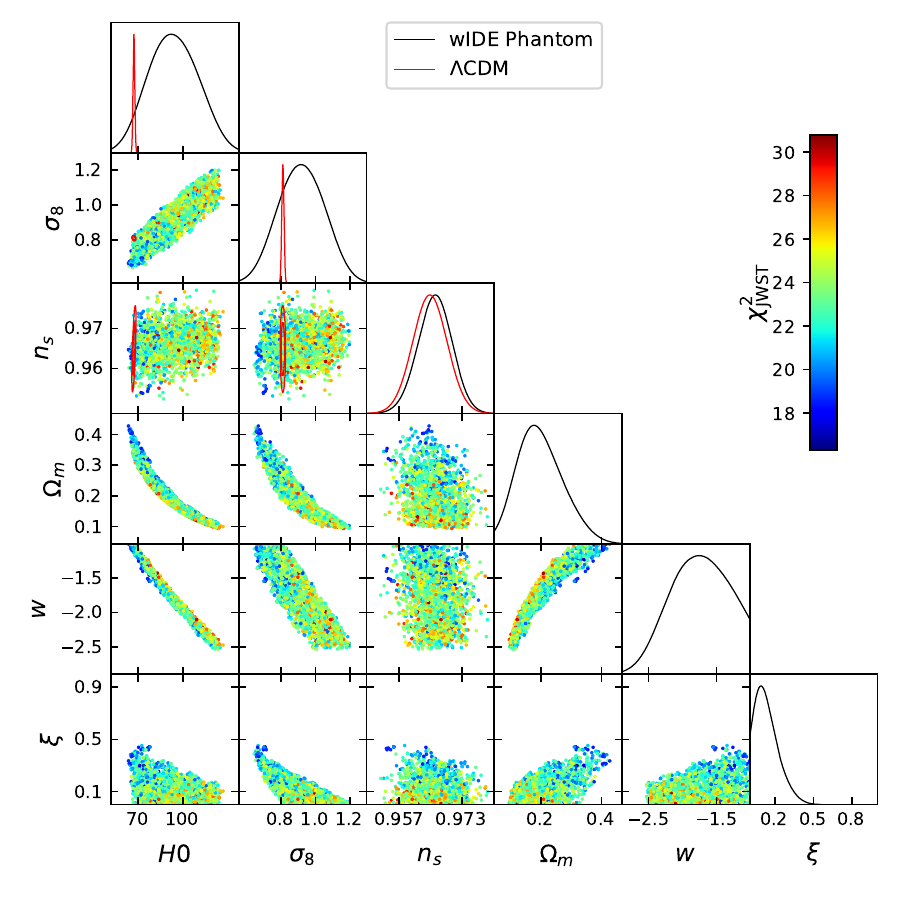}
	\caption{Triangular plot showing the distribution of points and the correlations among the most relevant parameters of $w$IDE, when $w$ is free to vary in the phantom region $w< -1$. The color map refers to the value of $\chi^2_{\rm JWST}$ so that the color pattern in the figure represents the direction towards which one needs to move in the parameter space to improve the fit to JWST data.}
	\label{fig:pwIDE_LCDM}
\end{figure*}

The best-fit values of cosmological parameters for the $w$IDE model with $w > -1$ (i.e., confined to the quintessence regime) are displayed in \autoref{tab:nwIDE}. Qualitatively, these results mirror those previously obtained for $w = -1$. There are no differences between the high and low redshift bins, and there is no preference for $\xi \neq 0$ from \textit{JWST-CEERS} data. The behaviors of the parameters depicted in \autoref{fig:nwIDE_LCDM} clearly indicate that introducing a coupling while leaving $w$ free to vary does not improve the fit to \textit{JWST-CEERS} observations. Once more, the reason behind this phenomenon is the decrease in matter density resulting from the energy flow within the interacting model.

The situation becomes somewhat more intricate when we turn to the case where the dark energy equation of state is confined to the phantom regime $w<-1$. In this case, the best-fit values of parameters are summarized in \autoref{tab:pwIDE} for the usual combinations of datasets. When considering only the best-fit values from the CMB, we find the well-documented Planck preference for a phantom equation of state~\cite{Yang:2021flj,Escamilla:2023oce}, with the best-fit value reading $w=-2.04$. Given the well-known degeneracy between the effects produced by a phantom $w$ and increasing the present-day expansion rate of the Universe, interacting phantom models can provide a much larger value of $H_0$~\cite{Yang:2022csz} and, in fact, we obtain a best-fit value $H_0=103.8$ km/s/Mpc. This essentially indicates that, without including datasets at lower redshifts, breaking this line of degeneracy proves challenging and values of $H_0$ in line with those measured by SH0ES collaboration can always be reintroduced by considering a sufficiently phantom dark energy component. In addition, due to a combination of correlations among different parameters such as $H_0$, $w$, and $\Omega_m$, we observe that for the latter parameter, the best-fit value reads $\Omega_m=0.139$ and is compensated by a substantial increase in $\sigma_8=1.026$. In light of these effects on parameters governing the matter clustering in the Universe, the question of whether (and to what extent) a phantom $w$IDE model could effectively contribute to explaining the anomalies observed in JWST remains a topic of debate. Looking at the brighter side, when we consider the fit to \textit{JWST-CEERS} data, we observed that both for the high ($\chi^2_{\rm JWST-CEERS}=16.3$) and low ($\chi^2_{\rm JWST-CEERS}=10.8$) redshift bins, $\chi^2_{\rm JWST-CEERS}$ slightly improves compared to the standard cosmological model. Furthermore, the \textit{JWST-CEERS} data seem to point toward a non-zero coupling $\xi\sim 0.2 - 0.3$, which is substantially higher than that preferred by the Planck data (whose best-fit value is $\xi\simeq0.05$). The reason underlying this preference is that to ensure the stability of the perturbations, $\xi$ is now required to be positive. This results in a shift of the energy flow from the dark energy sector to the dark matter one. Consequently, increasing the value of the coupling means injecting more power into the matter sector, facilitating structure formation, and improving the \textit{JWST-CEERS} fit. However, looking at \autoref{fig:pwIDE_LCDM} -- where, as usual, we plot the correlations among different parameters together with a color-map representing the value of $\chi^2_{\rm JWST-CEERS}$ -- it becomes really difficult to identify a color pattern representing the direction in which we need to move in the parameter space to improve the fit to \textit{JWST-CEERS} data. Despite this, with a good degree of imagination, we can speculate that by moving towards larger values of the coupling $\xi\to 0.4$ (corresponding to other values of $\Omega_m\to 0.4$), the value of $\chi^2_{\rm JWST-CEERS}$ seems to undergo a general improvement for the highlighted reasons, and its value becomes as good as the $\Lambda$CDM one. Interestingly, looking again at \autoref{fig:pwIDE_LCDM}, we note that the value of the Hubble rate corresponding to such a coupling is $H_0\sim 70$ km/s/Mpc, i.e., close to the result provided by SH0ES collaboration.

Therefore, in the context of IDE cosmology, the only potential scenario in which we can simultaneously slightly improve the agreement between the model predictions and JWST data while obtaining values of $H_0$ in line with local distance ladder measurements seems to involve considering a phantom component of the equation of state of dark energy. That being said, the ability of this model to address these two issues remains somewhat limited, above all when compared to the competing EDE solutions discussed in the previous section.

\section{Discussion and Conclusions}
\label{sec:conclusions}

Cosmological data have entered an era of high precision. Advanced telescopes with higher sensitivity are unveiling measurements at previously uncharted cosmological scales and epochs. However, this is a double-edged sword: while high-precision parameter extraction became achievable, a range of unexplained anomalies emerged, some of them growing in significance rather than diluting as the error bars decreased.

The most recent anomaly involves the JWST observations of a population of very high-mass galaxies at previously unexplored redshifts of $z\sim 7-10$. This suggests a higher cumulative stellar mass density in the redshift range $7 < z < 11$ than predicted by the standard $\Lambda$CDM model of cosmology. Assuming no systematic issue behind these preliminary findings, one might wonder whether these emerging anomalies could be somehow connected to other long-standing cosmological puzzles, such as the Hubble tension. All these issues might hint at a shared limitation in our current comprehension of the Universe, motivating the need to consider alternative theoretical scenarios.

In this paper, taking the four independent JWST datasets at face value, we first explore the correlation between the JWST likelihood and the fundamental six-$\Lambda$CDM parameters. We argue that, from a phenomenological standpoint, models with a larger matter component $\Omega_m$, a higher amplitude of primordial inflationary fluctuations $A_s$ and a bigger scalar spectral index $n_s$ are able to predict larger cumulative stellar mass densities, thus providing a better fit to JWST data. As a next step, we notice that part of this phenomenology is very common in models featuring new physics in the dark energy sector (both at early and late times) that have been recently proposed as possible solutions to the Hubble tension. In such models the behavior of the dark energy fluid can also influence the growth of structure, potentially leading to the formation of more massive galaxies able to account for the preliminary JWST measurements. Inspired by this idea, we explore whether in the context of Early Dark Energy or Interacting Dark Energy, the JWST findings could be explained, or at least, if within these scenarios, the fit to JWST observations is improved with respect to the one in the minimal $\Lambda$CDM framework.

We find that EDE, which leads to an increased $\Omega_m$ and $n_s$ concurrent with a rising EDE fraction $f_{\rm EDE}$, constitutes an excellent candidate. Not only we can improve the agreement between the theoretical predictions of the model and the JWST data (i.e., the minimum value of $\chi^2_{\rm JWST}$ obtained for both the high and low redshift bins is significantly better than what we get in $\Lambda$CDM), but to achieve this, we move through the parameter space in the same direction needed to solve the Hubble tension. This underscores that it is indeed possible to address both issues within the same theoretical framework.

Conversely, $w$IDE models featuring a dark energy equation of state $w \geq -1$ are generally disfavored from JWST, despite yielding higher values of matter clustering parameter $\sigma_8$. This is due to the energy flow from the dark matter sector to the dark energy one, implying a smaller $\Omega_m$. On the other hand, when the equation of state is confined to the phantom regime $w < -1$, the situation becomes somewhat more intricate. Whether, and to what extent, a phantom $w$IDE model could effectively contribute to explaining the anomalies observed in JWST findings remains an open question. The energy-momentum dynamics and parameter degeneracies can lead to a significant increase in the matter component, which in turn can slightly improve the agreement between the model predictions and the JWST data while also yielding values of $H_0$ in line with local distance ladder measurements. However, the ability of this model to address these two issues simultaneously remains limited compared to EDE. 

We conclude by pointing out that other studies have recently investigated dark energy and dark matter alternatives to reconcile the JWST anomalies. Just to mention a few appealing possibilities, in Ref.~\cite{Menci:2022wia} it was argued that adjusting the dark energy sector to accommodate a dynamic equation of state can be compatible with existing observations. However, this would require considering somewhat exotic dynamical dark energy models that necessitate specific configurations of dark energy parameters. Always in relation to the dark energy sector of the cosmological model, in Refs.~\cite{Menci:2024rbq,Adil:2023ara}, it was argued that an evolving DE component with positive energy density on top of a negative cosmological constant can be consistent with JWST observations. Other different yet interesting possibilities explored to approach the structure formation dilemma involve considering extensions related to particle physics, such as Fuzzy Dark Matter models consisting of ultra-light axions~\cite{Gong:2022qjx} or Warm Dark Matter~\cite{Dayal:2023nwi,Maio:2022lzg}. Axion models have the potential to mitigate the formation of smaller structures, aligning JWST's stellar mass density data with the CMB optical depth near $z \approx 8$. Instead, Warm Dark Matter models introduce a structural formation delay, predicting fewer low-mass systems at higher redshifts. However, it is worth noting that up until now, observations do not conflict with predictions from either cold dark matter or Warm Dark Matter theories for particles heavier than $2$ keV~\cite{Dayal:2023nwi}. Finally, another category of solutions competitive with those examined in this study frequently proposes new physics in the gravitational sector. For instance, if long-range attractive forces stronger than gravity are realized in nature, the formation of cosmic halos could begin during the radiation-dominated era, providing a possible explanation for the excess of heavy galaxies observed by JWST.

Future JWST data will either reinforce or diminish the need for exploring physics beyond the established $\Lambda$CDM scenario.

\acknowledgments 
MF, R and AM thank TASP, iniziativa specifica INFN for financial support.
EDV is supported by a Royal Society Dorothy Hodgkin Research Fellowship.
The work of AM was supported by the research grant number 2022E2J4RK "PANTHEON: Perspectives in Astroparticle and Neutrino THEory with Old and New messengers" under the program PRIN 2022 funded by the Italian Ministero dell’Universit\`a e della Ricerca (MUR). RCN thanks the financial support from the Conselho Nacional de Desenvolvimento Cient\'{i}fico e Tecnologico (CNPq, National Council for Scientific and Technological Development) under the project No. 304306/2022-3, and the Fundação de Amparo à pesquisa do Estado do RS (FAPERGS, Research Support Foundation of the State of RS) for partial financial support under the project No. 23/2551-0000848-3. This article is based upon work from COST Action CA21136 Addressing observational tensions in cosmology with systematics and fundamental physics (CosmoVerse) supported by COST (European Cooperation in Science and Technology). We acknowledge IT Services at The University of Sheffield for the provision of services for High Performance Computing. The work of OM is supported by the Spanish  grant PID2020-113644GB-I00 and by the European Union’s Framework Program for Research and Innovation Horizon 2020 (2014–2020) under grant H2020-MSCA-ITN-2019/860881-HIDDeN and SE project ASYMMETRY (HORIZON-MSCA-2021-SE-01/101086085-ASYMMETRY) and well as by the Generalitat Valenciana grant CIPROM/2022/69.

\appendix
\section{JWST-FRESCO survey uncertainty approximation}
\label{app:FRESCO}
In this appendix, we provide additional details about the methodology we used to derive the uncertainty on the cumulative comoving number density of dark matter halos $\Delta \log n$ for \textit{JWST-FRESCO}. We use the same statistical methodology adopted in Refs.~\cite{Menci:2022wia, Menci:2024rbq} and introduced in Refs.~\cite{1986ApJ...303..336G,Ebeling:2003tf}. At its core, the methodology relies on approximations to the exact Poissonian confidence limits for small numbers of observed events (that in our case is $3$). More quantitatively, we approximate the true Poissonian upper limit, by means of Eq.(4) of Ref.~\cite{Ebeling:2003tf}, that, for 3 events, reads
\begin{equation}
\Delta \log n_{\rm upper} = 4 \left[ \frac{35}{36} + \frac{S}{6} + 4^{c(S)} b(S)\right]^{3},
\end{equation}
where we fix $S\simeq 1.645$ which corresponds to choosing a 95\%CL interval uncertainty,\footnote{For Gaussian statistics (i.e. a normal probability distribution) the desired CL is related to $S$ by $$\text{CL}(S) = \frac{1}{\sqrt{2\pi}} \int_{-\infty}^{S} e^{-t^2 /2} \, dt. $$} see also the third column of Tab. 3 in Ref~\cite{1986ApJ...303..336G}. Notice also that, once $S$ is fixed, $c(S)$ and $b(S)$ are numerical coefficients that can be easily calculated by using Eqs.(6)-(7) in Ref.~\cite{Ebeling:2003tf}. Similarly, adopting Eq.(11) of Ref.~\cite{Ebeling:2003tf}, we estimate the lower limit on the error bar as:
\begin{equation}
\Delta  \log  n_{\rm lower} = 4 \left[\frac{26}{27} + \frac{S}{3\sqrt{3}} + 3^{\gamma(S)} \beta(S) + \delta(S) \sin\left(\frac{10\pi}{13}\right)\right]^3
\end{equation}
where, as usual, $S\simeq 1.645$ and $\beta(S)$, $\gamma(S)$ and $\delta(S)$ are given by Eqs.(9), (10) and (12) of Ref.~\cite{Ebeling:2003tf}, respectively. We stress that this statistical methodology, while accurate, necessarily introduces an additional layer of approximation. For this reason, we remain conservative proving the uncertainties on $\log{n(>M_{\rm halo})}$ at 95\% CL (corresponding to fixing $S \simeq 1.645$).

\bibliographystyle{JHEP}
\bibliography{Bibliography}

\end{document}